\documentclass[secnumarabic, twocolumn, graphics,floatfix, superscriptaddress,tightenlines, aps, prb]{revtex4-2}
\usepackage[dvips]{graphicx}
\usepackage{amssymb,amsmath}
\usepackage{dcolumn}
\usepackage{bm}
\usepackage{color}
\usepackage[colorlinks=true, allcolors=blue]{hyperref}
\usepackage{multirow}

\begin{document}

\title{Resonant optical cooling of nuclear spins in case of strong Knight field of photoexcited electrons }

\author{K.~V.~Kavokin}
\affiliation{Spin Optics Laboratory, St. Petersburg State University, 198504 St. Petersburg, Russia}

\begin{abstract}

Resonant cooling of the nuclear spin system of a semiconductor by spin-polarized charge carriers under pumping with helicity-modulated polarized light is considered theoretically. It is shown that in the case of strong Knight field of charge carriers, exceeding local fields of the dipole-dipole interaction of nuclear spins, the Overhauser field arising as a result of resonant cooling can considerably modify the overall shape of magnetic-field dependences of charge carrier spin polarization, experimentally observed as the Hanle effect.

\end{abstract}

\pacs{} \maketitle

\section{Introduction}
\label{sec:intro}
Resonant cooling of nuclear spins is observed under conditions of optical orientation of charge carrier spins by light with alternating circular polarization, which oscillates at a frequency close to the NMR frequency in an applied constant magnetic field perpendicular to the wave vector of the exciting light \cite{meier1984optical}. It usually manifests itself as a small feature on the magnetic depolarization curve of photoluminescence (the Hanle effect) \cite{meier1984optical}; observations of resonant cooling via the Kerr effect are also known \cite{Zhukov all-optical 2014}. This feature is due to the emergence of the Overhauser field of spin-polarized nuclei, which is added to or subtracted from the external field. Qualitatively, resonant cooling is explained by a combination of resonant pumping of the nuclear spin component rotating in an external field and the effect of the oscillating Knight field of electrons on it. Under these conditions, an energy flow occurs into or out of the nuclear spin system, leading to a shift in the energy distribution of nuclear spins toward the upper or lower portion of their energy spectrum, which is equivalent to cooling in the range of negative or positive spin temperatures, respectively. The increased magnetic susceptibility of cooled nuclear spins leads to their partial orientation along the external field. The sign of the nuclear spin temperature and, accordingly, the sign of the Overhauser field changes during resonance due to a change in the phase difference between the Knight field and the rotating nuclear spin. Therefore, the feature on the Hanle curve caused by resonant cooling has the form of a dispersion contour. Resonant cooling is quantitatively described using the formalism of the Provotorov equations \cite{MerkulovTkachuk} or the model of nuclear spin cooling in a rotating frame \cite{Kalevich Fleisher rotating}. As research into spin phenomena in new classes of materials (e.g., perovskites \cite{KoturBazhin}) and nanostructures progresses, a situation may arise where the Knight field of photoexcited electrons, typically comparable in magnitude to the local fields of the dipole-dipole interaction of neighboring nuclei, proves much stronger than these fields. In this case, one can expect a significant change in the shape of the Hanle curves under resonant cooling conditions, which is not described by existing theories. In this paper, we attempt to construct a theory of resonant cooling of nuclear spins at large (compared to local fields) amplitudes of the oscillating Knight field of photoexcited charge carriers. A modified rotating frame method is used for this purpose.

\begin{figure}
\centering
\includegraphics[width=1\linewidth]{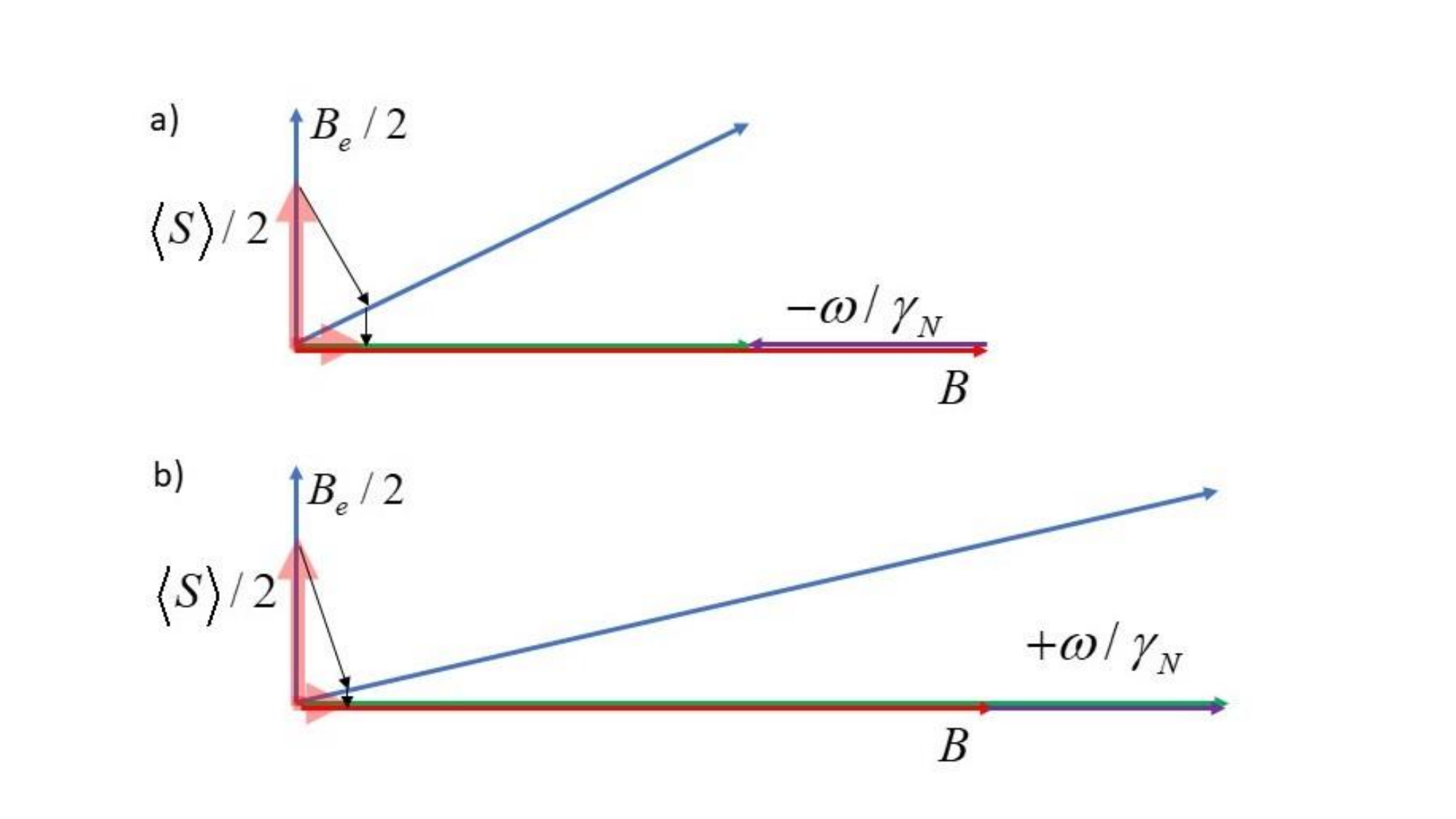}
\caption{A sketch of effective fields and spin components in the two rotating frames: a) the frame rotating along with the Larmor precession of nuclear spins and b) the frame rotating opposite to the Larmor precession. The generated average spin is projected onto the total field in the rotating frame, and then onto the constant effective field in the laboratory frame.}
\label{fig:1}
\end{figure}

An oscillating Knight field with frequency $\omega$ and amplitude $B_e$, which in the Hanle geometry is perpendicular to the external field, can be decomposed into two components rotating in opposite directions with amplitudes $B_e/2$. In each of the two coordinate systems rotating synchronously with these components, the nuclei experience a total field
\begin{equation}
\vec{B}_{\pm}^{*} = (B_e/2)\,\vec{e}_{z \pm} + (B \pm \omega/\gamma_N)\,\vec{e}_x
\label{eq:1}
\end{equation}
where $B$ is the external field directed along the $X$ axis in the laboratory frame.

Flip-flop transitions lead to a flow of non-equilibrium spin into the nuclear system. The vector of the electron spin generation rate is parallel to the Knight field and oscillates synchronously with it. It can also be decomposed into two rotating components. In each of the rotating frames there exists a stationary component of spin flow, parallel to a stationary Knight field. As a result, in each rotating frame dynamic nuclear polarization (DNP) occurs in the total field at a rate proportional to the projection of the generation rate onto the total field. The DNP rate along the $X$ axis, common to both rotating frames and the laboratory frame, is equal to the sum of the DNP rates along the $X$ axis in the two rotating frames (see Fig.~\ref{fig:1}). The components of spin flow and Knight field, oscillating in the rotating frames at double modulation frequency, do not contribute to the dynamic polarization since this frequency falls well outside the frequency range of the nuclear spin correlator \cite{meier1984optical}.

The steady-state nuclear polarization along the $X$ axis is determined by the balance between dynamic nuclear polarization (DNP) and spin-lattice relaxation induced both by interaction with electrons and by other mechanisms. Taking into account generation and relaxation leads to the following balance equation for the nuclear field along the $X$ axis:
\begin{equation}
\begin{aligned}
\frac{B_N}{T_1} &= \frac{1}{T_{1e}}\, b_N b_e \langle S \rangle^2 \cdot \\
&\cdot \Bigg[
\frac{B - \omega/\gamma_N}{b_e^2 \langle S \rangle^2 + (B - \omega/\gamma_N)^2}
+
\frac{B + \omega/\gamma_N}{b_e^2 \langle S \rangle^2 + (B + \omega/\gamma_N)^2}
\Bigg]
\end{aligned}
\label{eq:2}
\end{equation}
where
\begin{equation}
\langle S \rangle = \frac{\langle S_0 \rangle}{\sqrt{1 + (B + B_N)^2 / B_{1/2}^2}}
\label{eq:3}
\end{equation}

\begin{figure}
\centering
\includegraphics[width=1\linewidth]{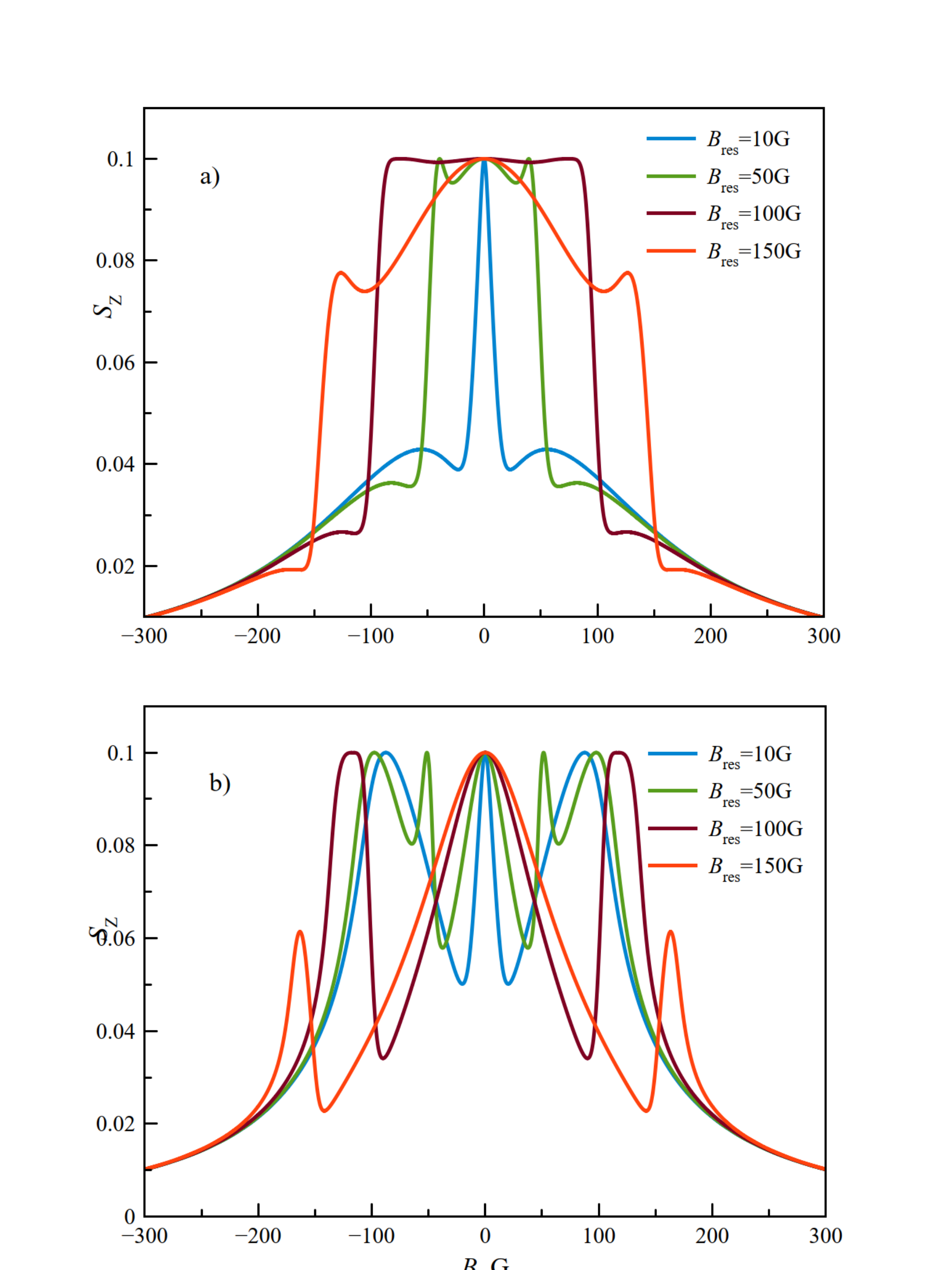}
\caption{Examples of Hanle curves under resonant cooling of nuclear spins for several modulation frequencies corresponding to resonance fields $B_{res}$ for the following parameters of the electron-nuclear spin system: $b_e=200G$, $b_N\frac{T_1}{T_{1e}}=2000G$, $S_0=0.1$, $B_{1/2}=100G$. a) negative g-factor of charge carriers, b) positive g-factor of charge carriers (the nuclear magnetic moment is positive for both cases) }
\label{fig:2}
\end{figure}

is the amplitude of the oscillating average electron spin under Hanle effect conditions. Here $b_e$ and $b_N$ are the Knight and Overhauser fields at full polarization of electron and nuclear spins, respectively, and $T_1$ and $T_{1e}$ are the characteristic times describing nuclear spin-lattice relaxation and dynamic nuclear polarization by electrons. In the absence of other relaxation mechanisms one has $T_1 = T_{1e}$. A numerical solution of the nonlinear Eq.~\eqref{eq:2} yields the nuclear field $B_N$ as a function of the external field $B$, which allows calculation of the Hanle curve shape. Examples of Hanle curves for several modulation frequencies are shown in Fig.~\ref{fig:2}. The sign of the right-hand side of Eq.~\eqref{eq:2} is determined by that of the product $b_N b_e$, that is, eventually by signs of magnetic moments of nuclei and charge carriers. This fact allows one to determine the sign of the charge carrier g-factor in the specific semiconductor structure from the shape of the resonance-cooling feature on the Hanle curve \cite{Kalevich Fleisher g-factor sign}. In case of strong Knight field, the entire shape of the Hanle curve becomes distinctively different for positive vs negative carrier g-factors.

\end{document}